1# CRS-FL: Conditional Random Sampling for Communication-Efficient and Privacy-Preserving Federated Learning

Jianhua Wang, Xiaolin Chang, Jelena Mišić, Vojislav B. Mišić, Lin Li, and Yingying Yao*Abstract*—Federated Learning (FL), a privacy-oriented distributed ML paradigm, is being gaining great interest in Internet of Things because of its capability to protect participants' data privacy. Studies have been conducted to address challenges existing in standard FL, including communication efficiency and privacy-preserving. But they cannot achieve the goal of making a tradeoff between communication efficiency and model accuracy while guaranteeing privacy.

This paper proposes a *C*onditional *R*andom *S*ampling (CRS) method and implements it into the standard FL settings (CRS-FL) to tackle the above-mentioned challenges. CRS explores a stochastic coefficient based on Poisson sampling to achieve a higher probability of obtaining zero-gradient unbiasedly, and then decreases the communication overhead effectively without model accuracy degradation. Moreover, we dig out the relaxation Local Differential Privacy (LDP) guarantee conditions of CRS theoretically. Extensive experiment results indicate that (1) in communication efficiency, CRS-FL performs better than the existing methods in metric accuracy per transmission byte without model accuracy reduction in more than 7% sampling ratio (# sampling size / # model size); (2) in privacy-preserving, CRS-FL achieves no accuracy reduction compared with LDP baselines while holding the efficiency, even exceeding them in model accuracy under more sampling ratio conditions.

*Index Terms*—Federated learning, differential privacy, Poisson sampling, communication efficiency## I. Introduction

Machine Learning (ML) plays crucial roles in data analytics [1], social network management [2], and service improvement [3]. Therefore, with the rapid development of Internet of Things (IoT) technology, an immense amount of data has been collected from remote intelligent devices, which conventional ML techniques with a centralized manner hardly process effectively. To overcome the afore-mentioned issue, distributed ML architecture is proposed [4]–[6], with computing parallelism, data parallelism, and model parallelism. However, statistical heterogeneity and participant privacy are not considered in standard distributed ML. Especially, privacy is a great concern in European laws, such as General Data Protection Regulation (GDPR) [7]. Hence, Federated Learning (FL) [8], as an advanced manner of distributed ML, has been proposed to address privacy-preserving and statistical heterogeneity issues.

In the FL training process, remote participants, such as IoT devices, laptops, and smartphones, join the training of the global model with a central server without sharing their private dataset. Each participant renews his local model via receiving the aggregated global update from the server. Nevertheless, there are existing at least the following two severe issues in FL.

- **Communication bottleneck**. As participants, plenty of resource-constrained IoT devices like smart cars and smartphones join the FL system. This incurs severe communication overhead between participants and the centralized server because of transmitting the whole global model with millions of parameters (e.g., ResNet50, AlexNet) in the model aggregation process of FL training [9]. In addition, due to the high flexibility and mobility of participants, the uploading task may be interrupted under a limited uplink channel. Therefore, the communication bottleneck arises [10].
- **Privacy leakage**. Though FL is proposed to address privacy leakage, several menaces still exist. Concretely, an adversary can extract crucial features from the transmitting global update and then reconstruct the private dataset to infer the member of the FL system [11] [12]. Participants are highly damaged by privacy leakage due to malicious behaviors and are unwilling to contribute their dataset to the federated model training.

There exist studies to overcome the above-mentioned issues. For promoting communication efficiency in FL, model compression methods [13] are proposed. As a kind of model compression method, sampling-based approaches [14]–[16] achieve a better tradeoff between communication efficiency and model accuracy than other compression approaches [17]–[20]. Privacy can be assured by Secure Multi-party Computation (SMC) and Differential Privacy (DP) [21]. Benefiting from requiring less extra computation resources, DP-based methods, including Central DP (CDP) and Local DP (LDP), obtain more interest in constraint resource FL environment [22], [23].

Recently, some studies have bridged communication efficiency and privacy-preserving [16], [24] in FL. However, they are not applicable in computation constraint IoT devices due to the used SMC-based method [24] or requiring extra encode and decode time without changeable sampling size [16].

Nonetheless, inspired by [16], [25]–[27], we observe that the sampling strategy has natural strength to bridge the communication efficiency [14] and privacy-preserving [28] in FL. Concretely, we can only reduce the sampling size when the privacy budget decreases without additive noise (less privacy budget means less privacy leakage). Concurrently, a less sampling size means more communication efficiency. Obviously, it satisfies our goals, which gain communication efficiency and privacy-preserving with the FL model accuracy tradeoff.

With this intuition, in this paper, we propose a *C*onditional *R*andom *S*ampling (CRS) method and implement it into the standard FL settings (CRS-FL) to remedy the problem of communication bottleneck and privacy leakage in FL. Concretely, based on Poisson sampling, CRS is designed with stochastic probabilities to filter the Top-K gradient in an unbiased way. Using a conditional random coefficient, CRS introduces more probability of selecting zero-gradient than [14] to reach less communication overhead. The specific application scenario is that, under the limited uplink environment, CRS-FL may actively obtain less communication overhead by sacrificing little accuracy. On the other hand, in privacy-preserving, we provide the relaxation [29], [30] LDP guarantee conditions of CRS.

Our contributions to this paper are as follows.

- We propose a *C*onditional *R*andom *S*ampling (CRS) method and deploy it in the standard FL settings (CRS-FL). Concretely, using a conditional random coefficient, CRS-FL drops the local update of participants in an unbiased manner and introduces probabilities to obtain a zero-gradient to reach the tradeoff between communication efficiency and model accuracy.
- We provide the privacy conditions of CRS to satisfy the relaxation Local Differential Privacy (LDP) guarantee and prove the unbiasedness of CRS. In the FL scenario, CRS-FL provides the same (or better) model accuracy as privacy-preserving baselines while maintaining communication efficiency.
- To the best of our knowledge, we are the first to consider the overall communication cost under the privacy guarantee without needing extra computation resources. Moreover, under some extremely limited uplink environments, the participants of CRS-FL could actively sacrifice little accuracy to upload fewer parameters for communication efficiency.

Extensive experiments are conducted under CIFAR-10/100 and FEMNIST datasets in the standard FL with FedAvg aggregation mechanism [8] and demonstrate as follows.

- **Communication-efficiency**. Under 7% sampling ratio (# sampling size / # model size, introduced in TABLE III), CRS-FL achieves the same level of accuracy as baselines, illustrated in Fig.4, Fig.5, and Fig.6. Meanwhile, CRS-FL reaches the best communication efficiency in metric Accuracy/OT (accuracy per transmission byte, introduced in Section V.B) stated in Fig.4 (d). CRS-FL achieves up to 9.4× improvement in Accuracy/OT under the CIFAR-100 dataset compared with the main competitor.
- **Privacy-preserving**. Under the $\epsilon \in [0.1, 1.0]$ and 0.7% sampling ratio, CRS-FL achieves the same level as FedAvg (LDP) with Laplace and is higher than the SOTA FedSGD-DPC [31]. Moreover, under the 7% sampling ratio, CRS-FL obtain the highest accuracy compared with the privacy-preserving baselines.
- Benefiting from the tradeoff between communication efficiency and model accuracy with a privacy guarantee, CRS-FL performs more efficiently than other privacy-preserving approaches while having a better privacy guarantee than communication efficiency methods.

The rest of our paper is organized as follows. Section II presents the preliminary of FL, DP, and sampling. Section III gives the problem statement of this paper, including scheme overview, design goals, and problem formulation. Section IV and Section V provide the scheme description and main results, respectively. Section VI gives the related work of communication efficiency and privacy-preserving. In the end, we conclude this paper in Section VII. The frequently used notations are described in TABLE I.

TABLE I    NOTATIONS FREQUENTLY USED IN SECTIONS II, III, AND IV

| Notation | Description |
| --- | --- |
| $\mathcal{L}(\cdot)$, $l(\cdot)$ | The loss function for the FL system and participants |
| $i$ | Device client $i$ |
| $\omega_\mathcal{G}$, $\omega_{\ell_i}$ | The global model and the local model of client $i$ |
| $m$ | Total number of device clients |
| $P_i$ | The personalized weight of client $i$ |
| $d$ | Parameter space |
| $\epsilon$ | The privacy budget |
| $\delta$ | The relaxation probability |
| $\mathcal{M}$ | The randomized mechanism |
| $\mathcal{D}$, $\mathcal{D}'$ | The neighboring dataset |
| $\mathcal{O}$ | The output of the $\mathcal{M}$ |
| $K$ | Sampling size |
| $p$ | Sampling probability |
| $x$, $x'$ | The neighboring relation private data |
| $\text{Perc}_S(\cdot)$ | The server perception of the device client's information |
| $B(\cdot)$ | Bernoulli Sampling |
| $\mathcal{P}$ | Poisson sampling |
| $(x_{i,j}, y_{i,j})$ | A pair of data and responding labels |
| $\Delta \mathcal{G}_{\ell i}^r$ | Local gradient update in the round $r$ of $i$ |
| $\mathcal{G}$ | Gradient |
| $\omega_\mathcal{G}^0$ | Initialized global model |
| $\eta$ | Learning rate |
| $R$ | Total number of round |
| $r$ | Round $r$ |
| $\mathcal{S}$ | CRS mechanism |
| $\tau$ | Threshold of top-k selection |
| $\alpha$ | Conditional random coefficient |
| $T$ | Priority value |
| $D_{KL}(\cdot)$ | KL-divergence |
| $\mathbb{E}(\cdot)$ | Condition expectation |
| $O(\cdot)$ | Asymptotic time or space complexity |
| $\Omega(\cdot)$ | Lower bound |

## II. PRELIMINARY

This section presents the preliminary of FL, DP, and sampling strategy.

*A. Federated Learning*

FL is pervasive with the strength of privacy-preserving and collaborating learning of participants. Generally, a standard FL system includes a set of distributed participants and a single cloud central server. The primary goal of FL is for participants are going to train an optimal global model without transmitting their private dataset. The overall optimization objective of the simplified FL process is as follows Eq (1).

$$\min_{W \subseteq \mathbb{R}^d} \mathcal{L}(\omega_{\mathcal{G}}) := \sum_{i \in [m]} \frac{P_i}{\sum_{i \in [m]} P_i} l(\omega_{\ell_i}) \quad (1)$$

where $\omega_{\mathcal{G}}$ denotes the global model and $\omega_{\ell_i}$ is the local model of participants $i$. The $P_i$ denotes the personalized weight of participants $i$, while $\mathcal{L}(\cdot)$ and $l(\cdot)$ are the loss function for the FL system and participants, respectively.

*B. Differential Privacy*

Differential Privacy (DP), first proposed by Dwork *et al.* [32], is widely used in data analytics and plays a crucial role in privacy-preserving FL. Now, we give the definition of standard DP and LDP.

**Definition 1.** $(\epsilon, \delta)$-**Differential Privacy.** *A randomized mechanism $\mathcal{M}$ satisfies $(\epsilon, \delta)$-DP w.r.t. $\epsilon, \delta \geq 0$ and neighboring dataset $\mathcal{D}$ and $\mathcal{D}'$ iff the output of $\mathcal{M}$ belonging to $\mathcal{O} \subseteq \text{Range}(\mathcal{M})$, and*

$$\Pr[\mathcal{M}(\mathcal{D}) \in \mathcal{O}] \leq e^{\epsilon} \Pr[\mathcal{M}(\mathcal{D}') \in \mathcal{O}] + \delta. \quad (2)$$

Notably, if $\delta$ satisfies $\delta = 0$, the mechanism $\mathcal{M}$ is $\epsilon$-DP, which is pure without relaxation probability [29]. In addition, $\text{Range}(\mathcal{M})$ denotes the set of all probable outputs $\mathcal{M}$.

The standard interpretation of the definition of $(\epsilon, \delta)$-DP is with the probability $1 - \delta$, the probability ratio of the output of mechanism $\mathcal{M}$ under two neighboring datasets $\mathcal{D}$ and $\mathcal{D}'$ is less than $e^{\epsilon}$. It demonstrates that little $\epsilon$ and $\delta$ means less probability of inferring private datasets, and privacy is guaranteed.

**Definition 2.** **Local Differential Privacy (LDP).** *A randomized mechanism $\mathcal{M}$ satisfies $(\epsilon, \delta)$-LDP w.r.t. neighboring relation private data $x$ and $x'$ of device client $D$, iff the output of $\mathcal{M}$ belonging to $\mathcal{O} \subseteq \text{Range}(\mathcal{S})$, and*

$$\Pr[\text{Perc}_{\mathcal{M}}(x) \in \mathcal{O}] \leq e^{\epsilon} \Pr[\text{Perc}_{\mathcal{M}}(x') \in \mathcal{O}] + \delta. \quad (3)$$

where $\text{Perc}_{\mathcal{M}}(\cdot)$ denotes the central server perception of the device client's information. Notably, the server usually is honest-but-curious, and maybe leads to inferring private information.

*C. Sampling*

Sampling is practical to decrease the communication cost [14], [33]. We provide a detailed definition of Bernoulli Sampling and Poisson Sampling of FL.

**Definition 3. Bernoulli Sampling.** *Assuming that sampling $K$ samples are in the finite population $d$, each sampling process is regarded as an independent **Bernoulli trial** to decide whether it is sampled under the same probability $p$. The probability of sampling $K$ samples is computed by **Binomial distribution**:*

$$\Pr[B(d, p) = K] = \binom{d}{K} p^K (1-p)^{d-K} \quad (4)$$

Notably, Bernoulli sampling is a particular form of Poisson sampling that allows the inclusion probability to be variable.

Now, we place the Poisson sampling strategy into our FL scenario.

**Definition 4. Poisson Sampling of FL.** *Assuming that the $d$-dimension satisfies $\mathcal{G}_{\ell} := \langle \mathcal{G}_{\ell 1}, ..., \mathcal{G}_{\ell j}, ..., \mathcal{G}_{\ell d} \rangle$, with variable probability $p$ in the different sampling process, the existing Poisson sampling mechanism $\mathcal{P}$ is to sample $\mathcal{G}_{\ell j}$ with the following:*

$$\mathcal{P}_j = \begin{cases} 1, & \text{w.p. } p \\ 0, & \text{w.p. } (1-p) \end{cases}, \quad (5)$$

*and the estimated value of the sampled:*

$$\mathcal{G}_{\ell j} = \begin{cases} \mathcal{G}_{\ell j} / p, & \text{w.p. } p \\ 0, & \text{w.p. } (1-p) \end{cases}, \quad (6)$$

where $\mathcal{P}_j = 1$ denotes $\mathcal{G}_{\ell j}$ is sampled and $\mathcal{P}_j = 0$ means that $\mathcal{G}_{\ell j}$ is not sampled.

III. PROBLEM STATEMENT

In this section, we first state the scheme overview. Then, we construct the problem formulation, and at last, we demonstrate the design goals of our system.

*A. Scheme Overview*

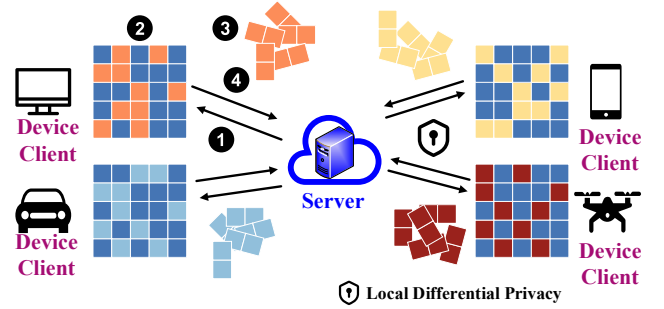

Fig.1. The overview of the CRS-FL.

Fig.1 presents the overview of the CRS-FL. There are including two entities: the central **Server** and remote **Device Clients**.

- **Server.** In the FL system, the server is responsible for broadcasting the global model, computing, and aggregating local updates. In general, the unprotected local update is likely to leak the private dataset information of participants [11]. Hence, there are existing servers with [34] /without [35] Trust Third Party (TTP). Notably, in our scenario, any TTP is not required. Moreover, we assume that the server is honest-but-curious, which means although the server obeys the standard FL protocol, it still is curious to eavesdrop and record the local model.
- **Device Client.** Device clients are the main participants of the FL system, including personal laptops, intelligent vehicles, mobile phones, unmanned aerial vehicles, etc. Each device client trains the global model utilizing their private dataset and uploads the local update to Server. We assume that different from fixed local data centers and large local servers with abundant computation and communication resources, device clients are resource-constrained participants. Nevertheless, the computation heavily depends on the size of the global model and local dataset. Accordingly, in this paper, we only focus on communication overhead reduction.

*B. Design Goals*

Due to the constrained resources and long-distance wireless transmission of IoT device clients in FL, the proposed CRS-FL cannot incur additional computation and communication overhead. Furthermore, the honest-but-curious server may cause severe privacy problems. Hence, the CRS-FL needs to meet the following design goals:

**Goal 1: Efficiency**. CRS-FL needs as far as possible to achieve succinct communication without compromising the model accuracy.

**Goal 2: Privacy**. The CRS-FL has to provide a privacy guarantee to prevent private information leakage.

**Goal 3: Unbiasedness**. The CRS-FL must satisfy an unbiased estimate of the device clients' local gradient to guarantee the unbiasedness of sampling results. It is worth mentioning that biased sampling will decrease the model accuracy with the increasing number of clients.

*C. Problem Formulation*

We have a set of $m$ device clients $D = \{D_1, ..., D_i, ..., D_m\}$ with their private dataset $\mathcal{D} = \{\mathcal{D}_1, ..., \mathcal{D}_i, ..., \mathcal{D}_m\}$, where $\mathcal{D}_i := \{(x_{i,j}, y_{i,j}) | j \in (1, n_i)\}$. The $n_i$ denotes the data number of the dataset $\mathcal{D}_i$, and $(x_{i,j}, y_{i,j})$ is a pair of data and responding label. We denote $\mathcal{L}(\omega_\mathcal{G}; (X, Y)): W \rightarrow \mathbb{R}$ as the loss function of the global model and $W \subset \mathbb{R}^d$ is the weight parameter space. Similarly, $l(\omega_{\ell_i}^r; (x_{i,j}, y_{i,j}))$ denotes the local loss function of $D_i$. Hence, using the FedAvg aggregation strategy, the optimization objective of the FL process with FedAvg [8] is stated as following Eq (7).

$$\min_{W \subseteq \mathbb{R}^d} \mathcal{L}(\omega_\mathcal{G}; (X, Y)) := \frac{1}{m} \sum_{i=1}^{m} l(\omega_{\ell_i}; (x_{i,j}, y_{i,j})) \quad (7)$$

Apart from the standard FL process, we utilize the CRS $\mathcal{S}$ to reach communication efficiency (**Goal 1**) and protect the privacy of device clients (**Goal 2**). Specifically, the $\epsilon$ and $\delta$ is the privacy budget and relaxation probability, and the sampling size $K \ll d$. In addition, to meet **Goal 3**, we should guarantee the unbiasedness of $\mathcal{S}$. The CRS-FL overall optimization objective is transferred as follows Eq (8).

$$\min_{W, \epsilon, \delta} \mathcal{L}(\omega_\mathcal{G}; (X, Y)) := \frac{1}{m} \sum_{i=1}^{m} l(\omega_{\ell_i}; (x_{i,j}, y_{i,j}); \mathcal{S}(\omega_{\ell_i}, \epsilon, \delta, K))$$

$$s.t. \ \delta \ll \frac{1}{d}, \ \delta \in [0, 1), \ \mathbb{E}(\omega) = \omega \quad (8)$$

IV. SCHEME DESCRIPTION

This section describes the CRS-FL and CRS in detail. Then, we present a privacy analysis of CRS-FL to demonstrate the privacy-preserving condition and unbiased proof.

*A. Description of CRS-FL*

Compared with the traditional FL training process, the CRS-FL add an additional step ❸, after that device clients train a global model by their own private dataset. Concretely, CRS-FL is described in **Algorithm 1**, and Fig. 1 illustrates the overview training process.

**STEP ❶: Broadcast global model.** At the beginning of training rounds, the server broadcasts the initialized global model $\omega_\mathcal{G}^0$ to each device client $D_i$.

**STEP ❷: Train the global model.** Only if it is the first training round, the device client $D_i$ obtain a local model $\omega_\ell^0$ using $\omega_\mathcal{G}^0$. On the other hand, $D_i$ receive global updates $\Delta \mathcal{G}_\mathcal{G}^r$ and compute the local model by $\omega_\ell^r \leftarrow \omega_\ell^{r-1} - \eta(\Delta \mathcal{G}_\mathcal{G}^r + \mathcal{G}_\mathcal{G}^{r-1})$. Subsequently, trained by local private dataset, $D_i$ computes loss $\mathcal{L}(\omega_\ell^r)$ and gets local gradient $\mathcal{G}_{\ell_i}^{r+1}$ of $D_i$. Then, the local update $\Delta \mathcal{G}_{\ell_i}^{r+1}$ is obtained by $\Delta \mathcal{G}_{\ell_i}^{r+1} \leftarrow \mathcal{G}_{\ell_i}^{r+1} - \Delta \mathcal{G}_{\ell_i}^r$.

**STEP ❸: Conduct the CRS.** Before transmitting the local update $\Delta \mathcal{G}_{\ell_i}^{r+1}$, the CRS is used, demonstrated in Section IV.B. At the end of the device client process, $D_i$ transmits the local update $\Delta \mathcal{G}_{\ell_i}^{r+1}$ to the server.

**STEP ❹: Aggregate local update.** The server aggregates the local update from each device client $D_i$ and conducts FedAvg to compute the global update by $\Delta \mathcal{G}_\mathcal{G}^r \leftarrow \frac{1}{m} \sum_{i=1}^{m} \Delta \mathcal{G}_{\ell_i}^r$. At last, the next iteration is going to begin.

*B. Description of CRS*

To effectively reduce the communication cost in constrained devices, the main intuition is to shorten the transmission bytes of local updates. In this paper, we proposed CRS to satisfy our design goals stated in **Algorithm 2**.

Respectively, in order to achieve **Goal 1**, we sample $K$-dimension vector from $d$-dimension local update $\Delta \mathcal{G}_{\ell_i}^r = \langle \Delta \|\mathcal{G}_{\ell 1}\|_n, \Delta \|\mathcal{G}_{\ell 2}\|_n, ..., \Delta \|\mathcal{G}_{\ell d}\|_n \rangle$ with the probability $p$. In fact, intuitively, it is a Bernoulli sampling problem, and probability is fixed by $p := K/d$.

However, to satisfy **Goal 2**, the probability of privacy leakage is unacceptable under the fixed $p := K/d$. We can imagine a scenario. The $D_1$ and $D_2$ own $d$-dimension vector and only 1-dimension are distinctive. Under the sampling size $K$, we have the probability of $K/d$ to leakage privacy of $D_2$, which is far beyond the ideal DP relaxation probability $\delta \ll 1/d$. Hence, we attempt to upgrade the sampling strategy from Bernoulli sampling to Poisson sampling, which has varying sampling probability of a single sampling process.

In the CRS, we assume that conditional random coefficient $\alpha_{\ell j}$ is arbitrary and satisfies $\alpha_{\ell j} \in [0, p)$, where $p$ is related to FL system privacy budget $\epsilon$ by $0 < p \leq 1 - e^{-\epsilon}$. Moreover, we introduce the randomness to each dimension $\Delta\|\mathcal{G}_{\ell d}\|_n$ of local update $\Delta\mathcal{G}_{\ell i}^r$ and compute priority value $T_{\ell j}$ by $T_{\ell j} \leftarrow \alpha_{\ell j} \cdot \|\mathcal{G}_{\ell j}\|^2$. Then we sort all of $T_{\ell j}$ by descending order and make Poisson sampling $\Pr[\mathcal{G}_{\ell j}|\Delta\mathcal{G}_{\ell i}^r] \leftarrow 1$ when $T_{\ell j} \geq \tau$ with $\tau \leftarrow \text{Top-k}(T_{\ell j})$. Up to now, we have finished the sampling process and satisfied **Goal 2**. Note that the proof of privacy guarantee is described in **Section IV.C**.

---

**Algorithm 1 CRS-FL**

**Input:** training round $R$, sampling value $K$, number of devices $m$, device client $D$, sampling privacy budget $\epsilon$, learning rate $\eta$

**Initialize:** global model $\omega_g^0$

1  **For** $r = 0, 1, 2, ..., R$ **do**
2  ################ Server-side Process ################
3  # **STEP 1: Broadcast global model**
4  **If** $r == 0$ **then:**
5      Broadcast $\omega_\mathcal{G}^0$ to $D_i$
6  **Else:**
7  # **STEP 4: Aggregate local update**
8      Receive local update $\Delta\mathcal{G}_{\ell i}^r$
9      $\Delta\mathcal{G}_\mathcal{G}^r \leftarrow \frac{1}{m}\sum_{i=1}^{m}\Delta\mathcal{G}_{\ell i}^r$  // FedAvg
10     Broadcast $\Delta\mathcal{G}_\mathcal{G}^r$ to $D_i$
11 **End if**
12 ################ Device-side Process ################
13 # **STEP 2: Train global model**
14 **If** $r == 0$ **then:**
15     $\omega_\ell^0 \leftarrow \omega_\mathcal{G}^0$
16 **Else:**
17     Receive global update $\Delta\mathcal{G}_\mathcal{G}^r$
18     $\omega_\ell^r \leftarrow \omega_\ell^{r-1} - \eta(\Delta\mathcal{G}_\mathcal{G}^r + \mathcal{G}_\mathcal{G}^{r-1})$
19 **End if**
20 **For** $D_i(i=1,...,m)$ **in parallel do**
21     $\mathcal{G}_{\ell i}^{r+1} \leftarrow \nabla\mathcal{L}(\omega_\ell^r)$
22     $\Delta\mathcal{G}_{\ell i}^{r+1} \leftarrow \mathcal{G}_{\ell i}^{r+1} - \Delta\mathcal{G}_{\ell i}^r$
23 # **STEP 3: Conduct CRS**
24     $\Delta\mathcal{G}_{\ell i}^{r+1} \leftarrow \mathcal{S}(R-1, \Delta\mathcal{G}_{\ell i}^{r+1}, K, \epsilon)$
25     Transmit $\Delta\mathcal{G}_{\ell i}^{r+1}$
26 **End for**
27 **End for**

---

It should be noted that an unbiased estimator of local updates during sampling is crucial. The overall model accuracy decreases remarkably in the communication efficiency approaches with biased estimators, such as Top-K, gradient clip, and gradient compression [14]. In this paper, to demonstrate the achievement of **Goal 3**, we provide detailed proof of the unbiasedness of the CRS in **Section IV.C**.

---

**Algorithm 2 CRS Process in Device-side**

**Input:** sampling value $K$, threshold $\tau$, local update $\Delta\mathcal{G}_{\ell i}^r$, $L_n$ norm, local vector $v$, vector dimension $d$, the sampling probability $p$

**Initialize:** $\Delta\mathcal{G}_{\ell i}^r = \langle\Delta\|\mathcal{G}_{\ell 1}\|_n, \Delta\|\mathcal{G}_{\ell 2}\|_n, ..., \Delta\|\mathcal{G}_{\ell d}\|_n\rangle$

1  **For** $j = 1, 2, ..., d$ **do**
2      Fixing sampling value $K$
3      $\alpha_{\ell j} \leftarrow \text{random}[0, p)$  # $p \in (0, 1-e^{-\epsilon}]$
4      $\mathcal{G}_{\ell j} := \Delta\|\mathcal{G}_{\ell j}\|_n$
5      $T_{\ell j} \leftarrow \alpha_{\ell j} \cdot \|\mathcal{G}_{\ell j}\|^2$  # Sampling with conditional random coefficient
6      $\tau \leftarrow \text{Top-k}(T_{\ell j})$  # Sort $T_{\ell j}$ by descending order
7      **If** $T_{\ell j} \geq \tau$ **then:**
8          $\Pr[\mathcal{G}_{\ell j}|\Delta\mathcal{G}_{\ell i}^r] \leftarrow 1$
9      **Else:**
10         $\Pr[\mathcal{G}_{\ell j}|\Delta\mathcal{G}_{\ell i}^r] \leftarrow 0$
11 **End if**
12 **End for**
13 **Return** $\Delta\mathcal{G}_{\ell i}^r$

---

### C. Privacy Analysis and Unbiased Proof

In this subsection, we will provide the relaxation LDP guarantee of our CRS (**Theorem 5**). In addition, from the perspective of system utility, we consider the privacy guarantee marginally meaningful with $\delta = O(1/d)$ [26], where $d$ denotes the size of the whole samples and the $O(\cdot)$ denotes the standard time complexity $O$ notation. Actually, we need to prove the relaxation probability $\delta$ satisfies $\delta \ll 1/d$ (**Lemma 6**). Finally, we present the unbiased guarantee to reflect our unbiased estimate of the local gradient (**Theorem 7**).

**Theorem 5.** *The CRS $\mathcal{S}$ obeys $(\epsilon, \delta)$-LDP in the neighboring data $x$ and $x'$ with the output of $\mathcal{S}$ belonging to $\mathcal{O}$, which $\delta$ satisfies $\delta \leq \exp\left\{-dD_{KL}\left(\frac{K}{d}\|p\right)\right\}$. In addition, we find the sample probability $p$ satisfy $0 < p \leq 1 - e^{-\epsilon}$ and the sample value $K$ satisfies $0 \leq K \leq d(1 - e^{-\epsilon} + pe^\epsilon)$.*

**Proof.** We divide our proof process into three parts, Part I is to demonstrate the necessary condition of the $\mathcal{S}$, Part II is to find the relaxation probability $\delta$, and Part III needs to seek the relationship between threshold $\tau$ and relaxation probability $\delta$.

**Part I.** We finish this part of the proof with inspiration from Theorem 2 of [27]. Specifically, we sample $K$ samples from $d$-dimension $\Delta\mathcal{G}_{\ell i}^r = \langle\Delta\|\mathcal{G}_{\ell 1}\|_n, \Delta\|\mathcal{G}_{\ell 2}\|_n, ..., \Delta\|\mathcal{G}_{\ell d}\|_n\rangle$. According to Binomial distribution with probability $p$ for sampling, we have $\Pr[X = K] = \binom{K}{d}p^K(1-p)^{d-K}$. Notably, $p$ in Poisson sampling is variable, but in a single-dimension sampling process, it is fixed.

According to **Definition 1**, assuming that $\mathcal{S}$ obeys $(\epsilon, \delta)$-LDP, notably only one data difference between $x$ and $x'$, we have $e^{-\epsilon} \leq \frac{\Pr[\text{Perc}_{\mathcal{S}}(x) \in \mathcal{O}]}{\Pr[\text{Perc}_{\mathcal{S}}(x') \in \mathcal{O}]} \leq e^{\epsilon}$, with probability $1 - \delta$.

Thus, if we sample $K$ samples from $d$, we can find

$$\frac{\Pr[X = K | \text{Perc}_{\mathcal{S}}(x) \in \mathcal{O}]}{\Pr[X = K | \text{Perc}_{\mathcal{S}}(x') \in \mathcal{O}]} = \frac{\binom{K}{d} p^K (1-p)^{d-K}}{\binom{K}{d-1} p^K (1-p)^{d-K-1}}$$

$$= \frac{\frac{d!}{K!(d-K)!} p^K (1-p)^{d-K}}{\frac{(d-1)!}{K!(d-K-1)!} p^K (1-p)^{d-K-1}} = \frac{d}{d-K}(1-p).$$

In this situation, if $\mathcal{S}$ obeys $(\epsilon, \delta)$-LDP, it must satisfy $e^{-\epsilon} \leq \frac{d}{d-K}(1-p) \leq e^{\epsilon}$.

**Part II.** For $\frac{d}{d-K}(1-p) \geq e^{-\epsilon}$, we could obtain the relationship between probability $p$ and $\epsilon$. Note that the sampled value $K \geq 0$ and when $K = 0$, we have $0 < p \leq 1 - e^{-\epsilon}$.

For $\frac{d}{d-K}(1-p) \leq e^{\epsilon}$, we could obtain the relationship between $k$ and $\epsilon$, and we have $K \leq d(1 - e^{-\epsilon} + pe^{-\epsilon})$. Now we compute the upper and lower bound of $K$, which satisfy $0 \leq K \leq d(1 - e^{-\epsilon} + pe^{-\epsilon})$.

**Part III.** This part is the main difference with [27]. Under the mechanism $\mathcal{S}$, we will sample the local update which satisfies $T_{\ell j} \geq \tau$. Recall that if $T_{\ell j} < \tau$, we set $\Delta \|\mathcal{G}_{\ell j}\|_n$ into zero. So intuitively, the "bad event" is when more than $K$ samples are sampled with probability $p$. Based on our designed mechanism, the sampled local update $\Delta \|\mathcal{G}_{\ell j}\|_n$ satisfies a Binomial distribution with $d$ trials and sampling probability $p$, namely $B(d, p)$. In this situation, we could find the relaxation probability $\delta$ satisfies $\Pr[B(d, p) > K]$.

Inspired by the Chernoff Bounding technique (Chapter 1 in [36]), we effectively connect probability and KL-divergence variables,

$$\Pr[B(d, p) > K] = \sum_{i \geq K} \binom{d}{i} p^i (1-p)^{d-i}$$
$$\leq \exp\left\{-dD_{KL}\left(\frac{K}{d} \| p\right)\right\}.$$

Thus, we have $\delta \leq \exp\left\{-dD_{KL}\left(\frac{K}{d} \| p\right)\right\}$, and $D_{KL}(\cdot)$ is the KL-divergence. ∎

**Lemma 6.** *The CRS $\mathcal{S}$ obeys $(\epsilon, \delta)$-LDP with $\delta \ll \frac{1}{d}$.*

**Proof.** We have found that the $\delta$ satisfies $\delta \leq \exp\left\{-dD_{KL}\left(\frac{K}{d} \| p\right)\right\}$. However, due to the Gibbs inequality $D_{KL}\left(\frac{K}{d} \| p\right) \geq 0$, and if and only if $\frac{K}{d} = p$, $D_{KL}\left(\frac{K}{d} \| p\right) = 0$. That means if $\frac{K}{d} \neq p$, $D_{KL} = D_{KL}\left(\frac{K}{d} \| p\right)$ satisfies $D_{KL} > 0$. In this situation, we need proof $\frac{K}{d} \neq p$.

According to the **Part II** of **Proof** of **Theorem 5**, $0 \leq K \leq d(1 - e^{-\epsilon} + pe^{-\epsilon})$, thus we have $\frac{K}{d} \leq 1 - e^{-\epsilon} + pe^{-\epsilon}$. If we assume that $\frac{K}{d} = p$, the privacy budget needs to satisfy $\epsilon = 0$, which is impossible in our mechanism $\mathcal{S}$.

So, we can prove that the relaxation probability $\delta$ satisfies $\delta \leq e^{-dD_{KL}} \ll O\left(\frac{1}{d}\right)$ when $d$ becomes infinity and obeys the privacy guarantee demand. ∎

**Theorem 7.** *The CRS $\mathcal{S}$ is unbiased, where the expectation satisfies $\mathbb{E}(\omega) = \omega$.*

**Proof.** We define $A := \{\alpha_{\ell 1}, \alpha_{\ell 2}, ..., \alpha_{\ell j}, ..., \alpha_{\ell d}\}$. In spiration from [14], notably, $\tau$ is the $K$-th $T_{\ell j}$, and we need to prove that $\mathbb{E}(\mathcal{G}|A) = \mathcal{G}$. Then we can further obtain $\mathbb{E}(\omega) = \omega$.

Based on the conditional expectation calculation principle of discrete random variables, we have

$$\mathbb{E}(\mathcal{G}|A)$$
$$= \mathbb{E}[\mathcal{G}|A = \alpha_j]$$
$$= \sum_j \mathcal{G}_{\ell j} \cdot \Pr[\mathcal{G} = \mathcal{G}_{\ell j} | A = \alpha_{\ell j}]$$
$$= \sum_j \mathcal{G}_{\ell j} \cdot \frac{\Pr[V = \mathcal{G}_{\ell j}, A = \alpha_{\ell j}]}{\Pr[A = \alpha_{\ell j}]}.$$

According to the definition of Poisson Sampling and **Theorem 5**, we get $\mathcal{G}_{\ell j} = \frac{\mathcal{G}_{\ell j}}{p_{\ell j}}$, notably $\alpha_{\ell j} \in [0, p_{\ell j})$ and $p_{\ell j} \in (0, 1 - e^{-\epsilon}]$. Furthermore,

$$= \frac{\Pr[\mathcal{G} = \mathcal{G}_{\ell j}, A = \alpha_{\ell j}]}{\Pr[A = \alpha_{\ell j}]}$$

$$= \frac{\Pr[\mathcal{G} = \mathcal{G}_{\ell j}] \cdot \Pr\left[A = \alpha_{\ell j} \geq \frac{\tau}{\mathcal{G}^2} \middle| \mathcal{G} = \mathcal{G}_{\ell j}\right]}{\Pr[A = \alpha_{\ell j}]}.$$

$$= \frac{1 \cdot p_{\ell j}}{1} = p_{\ell j}$$

In this situation, we have $\mathbb{E}(\mathcal{G}|A) = \sum \frac{\mathcal{G}_{\ell j}}{p_{\ell j}} \cdot p_{\ell j} = \mathcal{G}$. In this situation, we can find that each local update is unbiased, and we can further obtain $\mathbb{E}(\omega) = \omega$. ∎

## V. RESULTS AND ANALYSIS

In this section, we present the experimental settings. Then, evaluation results are demonstrated. Respectively, we provide the metric accuracy per overall transmission byte and disclose the performance which satisfies **Goal 1**, the metric accuracy with privacy-preserving to meet **Goal 2**, and the accuracy performance under various device client numbers to illustrate **Goal 3**.

*A. Experimental Settings*

We implement extensive experiments to evaluate CRS-FL. This subsection will present the experimental details, including datasets, models, baselines, and hyper-parameters.

*1) Datasets*
- **CIFAR-10.** CIFAR-10 [37] is made up of 10 classes of 32x32 images with three RGB channels and consists of 50,000 training samples and 10,000 testing samples.
- **CIFAR-100.** Similarly, CIFAR-100 [37] has the same size of images. However, it exists 100 classes with 600 samples (500 training samples and 100 test samples) in each class. In other words, CIFAR-100 is a more sophisticated dataset than CIFAR-10 to evaluate the FL system fairly.
- **Federated EMNIST.** Federated EMNIST (FEMNIST) [38] is an open-source federated dataset with over 3,500 device clients and 800,000 image samples. It provides hyper-parameters to determine the split data manner, such as i.i.d (independent identically distribution) and non-i.i.d, minimum samples of each device client, etc.

CIFAR-10 and CIFAR-100 are the famous benchmark in computer vision. In addition, FEMINIST has been proposed recently for appraising the performance of the FL scheme. To evaluate the CRS-FL with more realistic, we adopt non-i.i.d data split to simulate the real-world data distribution.

*2) Models*

**ResNet9.** ResNet9, with 6.5M parameters proposed by [39], is utilized to evaluate the performance of various approaches.

*3) Baselines*

Our CRS-FL aims to achieve efficient communication and privacy-preserving; hence the baselines we compared are classified by communication efficiency approaches and privacy methods. The asymptotic time complexity, asymptotic space complexity, unbiased estimator, communication approach, and privacy-preserving approach are concluded in TABLE II. It needs to state that CRS-FL has time complexity $O(K \log d)$ because of the top-k selection and lower bound space complexity $\Omega(d+0)$ with probability $p_\Omega = \left(1/\left(1-e^{-\epsilon}\right)\right)^K$, which is efficient and cost-friendly communication.

*(a) Comm. Efficiency*

**MinMax** [14]. MinMax adopts an unbiased Poisson-based sampling strategy to decrease the transmission overhead. They provide optimal MinMax with provable optimality and adaptive MinMax with near-optimal. Due to the tradeoff of efficiency and accuracy, adaptive MinMax is more realistic. In this paper, we adopt adaptive MinMax to compare. Notably, CRS-FL is derived from MinMax and expands the privacy guarantee of MinMax. In other words, MinMax is our significant comparison in *Comm. Efficiency*. We will provide a multi-dimension comparison with MinMax.

**GSpar** [15]. GSpar transmits a sparse gradient strategy to reduce the communication cost, which randomly drops out the stochastic gradient vectors and amplifies the remaining gradients to realize the unbiasedness of sparsified gradients.

**SparseSGD** [33]. SparseSGD provides the other Top-K strategy with a memory or feedback mechanism. Roughly speaking, they propose MEM-SGD update the largest changing magnitudes and reduce the non-zero entries in global model update and broadcast processing.

**Top-K** [40]. Top-K is a traditional efficient gradient compression method to transmit data. They select the top-k gradient and accumulate the rest gradient locally to avoid losing information. We adopt the accumulating strategy in experiments as well. However, the Top-K is a biased estimator of the local gradient, which may be impacted by increasing participants [14].

**FetchSGD** [17]. FetchSGD compresses model updates using a Count Sketch technique. They try to move momentum and error accumulation from clients to the central server to improve accuracy and decrease leaky gradient information.

**FedAvg** [8]. Benefiting from the simplicity of averaging the weights of participants, FedAvg is the most used baseline without privacy and communication efficiency strategy.

*(b) Privacy*

**FedSGD-DPC** [31]. FedSGD-DPC, as the SOTA proposed recently, investigates optimal numbers of queries and replies in FL with DP and attempts to maximize the final accuracy using randomly selecting participating clients under a uniform distribution manner to conduct local iterations.

**FedAvg (LDP).** As most studies do, we combine FedAvg with the Laplace mechanism, which is a pure $(\epsilon,0)$-LDP mechanism with Laplace noise, to construct the privacy baseline.

*4) Hyper-parameters*

As demonstrated in TABLE III, we present the used hyper-parameters of CIFAR and FEMNIST. Notably, FetchSGD is a count-sketch-based approach with a different sampling size $K$. The uploading bytes of each iteration round of FetchSGD are $\text{len}(rows) \times \text{len}(columns)$, the same as the sampling size in our scenario. Furthermore, due to the particularity of FEMNIST with 3597 users, we only conduct 1,200 iterations to speed up the experiments. Actually, the hyper-parameters are mostly the same as [14], [17], [31], which are the crucial comparisons of our paper.

### B. Evaluation Metrics

**Accuracy** and **CE Loss**. We use the classical metrics of accuracy and Cross-Entropy Loss (CE Loss) to evaluate the actual performance of the CRS-FL. Notably, due to the time-consuming of FEMNIST, we utilize Top 1 Accuracy (in training) instead of test accuracy.

**Overall Transmission** (OT). We compute the download transmission bytes and upload transmission bytes of overall training iterations per device client, and the detail equal is as follows Eq (9).

$$OT = (\#\text{download bytes} + \#\text{upload bytes})/m \quad (9)$$

**Accuracy/OT**. Assuming that the transferring speed rate of local updates from remote device clients to the central server is constant, the fewer bytes transmitted, the more efficiency. Plainly, we use Accuracy/OT to measure the accuracy improvement per byte and expose the efficiency of CRS-FL.

### C. Experimental Results of CRS-FL

In this subsection, we first present the accuracy scales in 2,400 iterations when the sampling size $K$ is 10k, 50k, 500k, and 3000k, respectively. It is worth explaining that the whole ResNet-9 owns almost 6.9M parameters, which means that we only upload less than half the size of ResNet-9 even though we use 3000k as the sampling size.

TABLE II  THE COMPARISON OF VARIOUS APPROACHES

| Approach | Based | Time Complexity † | Comm. Cost (OT) Upper Bound * | Comm. Cost (OT) Lower Bound ‡ | Unbiased Estimator § | Comm. Efficiency § | Privacy § |
|---|---|---|---|---|---|---|---|
| **CRS-FL (Ours)** | Sampling | $O(K\log d)$ | $O(d+K)$ | $\Omega(d+0)$ with $p_\Omega = (1/(1-e^{-\epsilon}))^k$ | ● | ● | ● |
| MinMax [14] | Sampling | $O(K\log d)$ | $O(d+K)$ | $\Omega(d+K)$ | ● | ● | ○ |
| GSpar [15] | Sampling | $O(d)$ | $O(d+K)$ | $\Omega(d+0)$ with $p_\Omega = (1-p)^k$ | ● | ● | ○ |
| SparseSGD [33] | Compression | $O(K\log d)$ | $O(K+d)$ | $\Omega(0+d)$ with $p_\Omega = (1-p)^k$ | ○ | ● | ○ |
| Top-K [40] | Compression | $O(d)$ | $O(d+K)$ | $\Omega(d+K)$ | ○ | ● | ○ |
| FetchSGD [17] | Sketch | $O(\log d)$ | $O(d+\log(dR/\delta)/c^2)$ | $\Omega(d+\log(dR/\delta)/c)$ | ● | ● | ○ |
| FedAvg [8] | - | $O(1)$ | $O(d+d)$ | - | - | ○ | ○ |
| FedSGD-DPC [31] | - | $O(d)$ | $O(d+d)$ | - | - | ○ | ● |
| FedAvg (Laplace LDP) | - | $O(d)$ | $O(d+d)$ | - | - | ○ | ● |

† *The Comm. Cost indicates the asymptotic time complexity. CRS-FL, MinMax, Top-K, and SparseSGD are top-k selection problems, while GSpar is a linear problem.*

* *The Comm. Cost (OT) Upper Bound indicates the upper bound of overall transmission bytes. Specifically, in asymptotic space complexity $O(\text{left}+\text{right})$ with $K \ll d$, **left** means downloading from sever and **right** means uploading from the client. Notably, $O(\log(dR/\delta)/c)$ [17] with little $\delta$ and $c \in (0,1)$.*

‡ *The Comm. Cost (OT) Lower Bound indicates the lower bound of overall transmission bytes. Notably, we assume that each gradient dimension is **non-zero**. MinMax has the same OT because MinMax has to upload the whole-size local update with the probability $p=1$ due to their sampling strategy. $p_\Omega$ denotes the probability of lower bounds.*

§ *The symbol ● indicates YES, while ○ is the opposite NO.*

TABLE III  THE HYPER-PARAMETERS OF EXPERIMENTS IN DIFFERENT DATASETS

| Dataset | Model | Data Type | Device Number | Sampling Size $K$ † | Iteration | DP Setting | Other Hyper-parameters |
|---|---|---|---|---|---|---|---|
| CIFAR-10 CIFAR-100 | ResNet-9 | Non-i.i.d | 200 in subsections C and D and 200, 500, and 800 in E. | [5, 10, 20, 30, 40, 50, 100, 200, 300, 400, 500, 650, 1300, 2000, 3000]$\times 10^3$, notably columns number [650, 1300, 2000, 3000] $\times 10^3$ for FetchSGD | 2400 | Laplace noise scale $\sigma = 1e-5$, Privacy budget $\epsilon \in [0.1, 1.0]$ | Learning Rate 0.01 with decay $1e-4$, local batch size 50 for CIFAR-10/100 and 30 for FEMNIST, FedAvg aggregation method |
| FEMNIST | | | 3597 | [500, 1000, 2000, 3000]$\times 10^3$, notably columns number [1000, 2000, 3000]$\times 10^3$ for FetchSGD | 1200 | | |

† *In CIFAR-10/100, the whole size of ResNet-9 is 6.9M bytes. Hence, [5, 50, 100, 200, 300, 400, 500, 650, 1300, 2000, 3000]$\times 10^3$ means [0.07%, 0.7%, 1%, 3%, 4%, 6%, 7%, 9%, 19%, 29%, 43%] sampling ratios.*

The precise lines are illustrated in Fig.2, and we evaluate CRS-FL in CIFAR-100 with the privacy budget satisfies $\epsilon \in [0.1, 1.0]$. Then we can observe that with the decreasing sampling size, the bound of accuracy scales with various $\epsilon$ is looser. In fact, tighter bounds imply less erratically with various $\epsilon$ and the accuracy tends to converge.

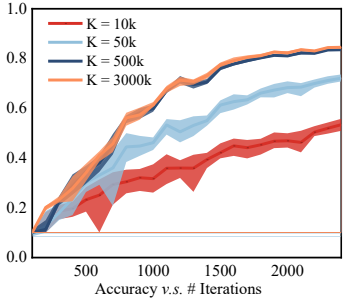

Fig.2. The accuracy *v.s.* iterations of CRS-FL in CIFAR-100 with $\epsilon \in [0.1, 1.0]$.

Moreover, to evaluate explicitly, we compare CRS-FL with the standard baseline FedAvg and FedAvg (LDP) in CIFAR-10 and CIFAR-100 with privacy budget $\epsilon = 1.0$, demonstrated in Fig.3. Specifically, the red dash lines are the accuracy of FedAvg. In contrast, dark blue dash lines are the CE loss of FedAvg because the accuracy and CE loss are not changed due to increasing sampling size $K$. In addition, the red arrows are the accuracy of FedAvg (LDP), while the dark blue arrows are the CE loss of FedAvg (LDP). Obviously, with the growth of sampling size $K$, the accuracy of our CRS-FL is nearly the uncompressed baseline FedAvg, after $K = 500k$ in particular. Similarly, the CE loss lines indicate the same trend.

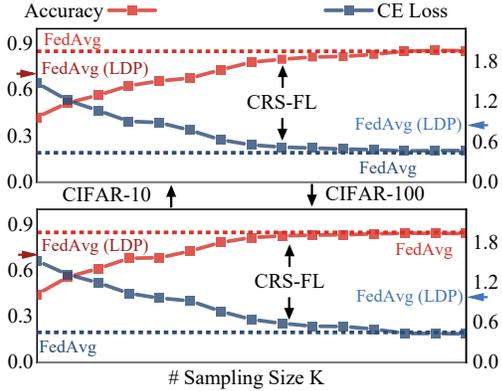

Fig.3. The accuracy and CE loss *v.s.* sampling size K with $\epsilon = 1.0$ in dataset CIFAR-10 and CIFAR-100.

### D. Comparisons of Comm. Efficiency

In this subsection, we implement extensive experiments in three datasets and compare our CRS-FL with the other five approaches to communication efficiency. Notably, the result of the baseline FedAvg is illustrated by a red line in each plot.

In the FEMNIST dataset stated in Fig.4, generally results we observe are similar to CIFAR. As described in Fig.4 (d), the Accuracy/OT of CRS-FL is higher than others. Therefore in $K = 3000k$, the FetchSGD achieves better performance, but they cannot reach the same accuracy level as FedAvg.

Generally speaking, the CRS-FL achieves the convergence level accuracy and CE loss with $K \geq 500k$ in CIFAR and $K \geq 2000k$ in FEMNIST. In addition, CRS-FL transmits the second least bytes in CIFAR and the least in FEMNIST. Moreover, the Accuracy/OT of CRS-FL is the second best in CIFAR and the best in FEMNIST, except $K = 3000k$. Notably, the first Accuracy/OT shows decreasing model accuracy with the growth of participants stated in Section V.E. With this intuition, CRS-FL can better reach the tradeoff between communication efficiency and accuracy performance.

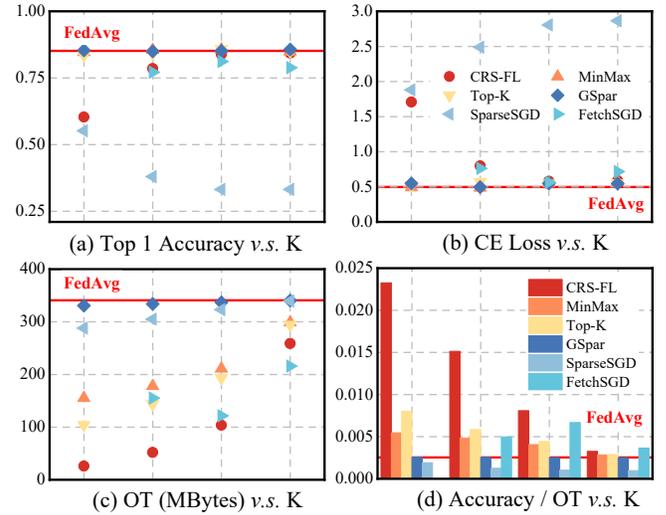

Fig.4. The result of four metrics in FEMNIST: (a) accuracy *v.s.* sampling size K; (b) CE loss *v.s.* sampling size K; (c) OT *v.s.* sampling size K; (d) accuracy / OT *v.s.* sampling size K.

### E. Comparisons of Unbiasedness

As we mentioned, Top-K is the biased estimator of the local gradient. However, the OT and Accuracy/OT results of Top-K are similar to CRS-FL under CIFAR. We will present the experimental results in CIFAR-100 with different device client numbers, 200, 500, and 800, respectively.

Fig.5 illustrates that from client numbers 200 to 500, the MinMax and CRS-FL, which are unbiased estimators, show a little accuracy drop. While from number 500 to 800, MinMax and CRS-FL nearly zero accuracy drop. However, with the increasing client numbers, the accuracy of Top-K drops remarkably, which is unacceptable.

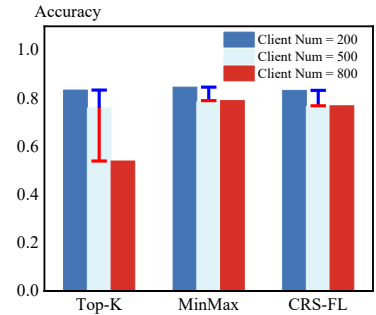

Fig.5. Various comparisons of different device client numbers in CIFAR-100.

*F. Comparisons of Privacy-preserving*

In this subsection, we implement the comparisons of privacy-preserving. We conduct experiments using the same hyper-parameters of FedSGD-DPC, which is open-source. Moreover, we have organized the experimental result plots as Fig.6. To show the results explicitly, we only provide the CRS-FL result with $K = 50k$ and $K = 500k$. As stated in Fig.6, in the left y-axis, the red circle is the accuracy with $\epsilon = 0.1$, and the cyan triangle is the accuracy with $\epsilon = 1.0$. In contrast, in the right y-axis, the red histogram denotes the Accuracy/OT with $\epsilon = 0.1$, and the blue histogram indicates the Accuracy/OT with $\epsilon = 1.0$.

Obviously, under the LDP privacy strategy, our CRS-FL with $K = 50k$ reaches the highest Accuracy/OT under three datasets, which means better communication efficiency. Furthermore, CRS-FL also achieves the same accuracy in $\epsilon = 1.0$ as FedAvg (LDP) and FedSGD-DPC under FEMNIST.

With the hyper-parameter K satisfies $K = 500k$, our CRS-FL reduces the communication efficiency (nonetheless, still more efficient than other competitors) but achieves the best accuracy under three datasets. In short, our CRS-FL realizes the better tradeoff with privacy, efficiency, and accuracy performance.

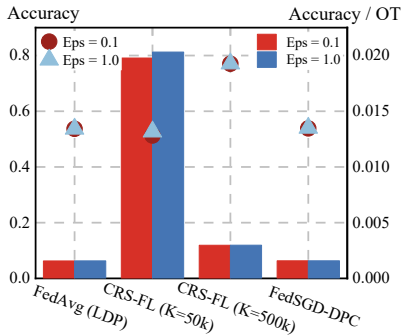

Fig.6. The accuracy and Accuracy/OT of privacy comparisons in FEMNIST.

## VI. RELATED WORK

This section presents the related work of communication efficiency, privacy-preserving, and both of them in FL.

**1) Communication Efficiency.** To address the issue of communication bottlenecks of FL, researchers concentrated on model compression studies to promote communication efficiency recently. The commonly used gradient compression approaches are classified [13] as sparsification-based [14], [15], [17], [33], [40], quantization-based [41], [42], knowledge distillation-based [43], and low-rank factorization-based [44]. Quantization-based approaches decrease the parameters in bit-width level instead of uploading directly, for example, changing float to integer, which significantly damages the performance of the system [18]. Knowledge distillation-based methods alleviate the communication burden by training a smaller student model in the server to imitate the teacher model in participants, which is trained mainly by the teacher model. However, extra training overhead and negative transfer exist due to the heterogeneity between public and private data [19]. Low-rank factorization-based methods focus on estimating the most informative parameters using the matrix factorization technique, but it cannot reduce communication cost-effectively [20]. The sparsification-based approaches, including compression-based [40], sketch-based [17], [45], and sampling-based [14], [15], regenerate stochastic gradients independently by unbiased or biased estimates of true gradients. However, compression-based and sketch-based methods fail to adjust the sparsification size dynamically. Moreover, the sketch-based methods have to negotiate in time to ensure the same sketch size. In addition, with the immense number of participants, a biased estimator (Top-K [40]) may decrease the accuracy of the FL system. In other words, the unbiased estimator is necessary for sparsification to retain the most gradient information of participants.

**2) Privacy-preserving.** In the privacy-preserving FL scheme, the mainstream methods are SMC-based and DP-based [21]. In SMC-based methods, the global model and local update are protected with cryptographic techniques, including homomorphic encryption and secret sharing. However, because of their extensive computation overhead, most SMC-based privacy-preserving approaches are hardly a tradeoff between efficiency and privacy [34]. In DP-based methods, there exist two categories, namely CDP and LDP. The CDP introduces the stochastic perturbation in the central server to protect the global model against service purchasers, while the LDP adds perturbation in the client to protect the local update. Due to the extra perturbation (Laplace noise, Gaussian noise, or other randomized mechanisms), the DP mechanism is subjected to the tradeoff between performance and privacy [46]. Compared with SMC-based, however, for the computation resource constraint participants (e.g., IoT devices), DP-based privacy-preserving strategies are more favored [13], [22], [23].

**3) Privacy-preserving and communication efficiency.** There are a few studies to address the above issues together. CLDP-SGD [24] is proposed using a shuffle DP mechanism for privacy guarantee and compressing gradient unbiasedly to reduce communication. However, an additional shuffle aggregator between participants and the server needs two more times remote communication in each training epoch, which is a needless consuming cost. Moreover, the shuffle mechanism still relies on Trust and Honest Third Party. Chen *et al.* [16] introduce the Poisson Binomial mechanism with Renyi DP guarantee into FL and show the same privacy-accuracy tradeoff as the Gaussian mechanism. However, they utilize an encoder in participants and a decoder in the server, which needs additional time complexity. In addition, the sampling probability is dominated by rescaling using several variables, which are not dynamically controlled by sampling size. According to those studies, since sampling can create uncertainty, we have found that it can offer a bridge between privacy and efficiency.

## VII. CONCLUSION

With the rapid development of the Internet of Things (IoT), IoT data is intense growth in the real world. As a widely used technique, Machine Learning (ML) plays a crucial role in data analytics, service promotion, and social network management. Federated Learning (FL), a distributed ML technique, is proposed to train a global model on the local side with privacy concerns. However, mobile smartphones and vehicles cannot provide abundant computation and communication resources. With the intuition of privacy and communication concerns, we propose a *C*onditional *R*andom *S*ampling (CRS) method in FL (CRS-FL) with a stochastic coefficient in an unbiased manner, which effectively reduces the communication overhead in uploading local updates from participants. Besides, we provide the LDP guarantee of CRS to protect the privacy of device clients. Extensive experiments illustrate that CRS-FL achieves the best communication efficiency via accuracy per transmission byte without accuracy reduction. Compared with privacy-preserving methods, CRS-FL shows the same accuracy level in 500k parameters sampled, even exceeding in more sampling size.